# Amorphous Selenium Mie Resonators for Infrared Meta-Optics

Danveer Singh[1†], Michal Poplinger[1†], Avraham Twitto[1], Rafi Snitkoff[1], Pilkhaz Nanikashvili[1], Ori Azolay[1], Adi Levi[1], Chen Stern[1], Gili Cohen Taguri[2], Asaf Albo[1,2], Doron Naveh[1,2*] and Tomer Lewi[1,2*]

[1]Faculty of Engineering, Bar-Ilan University, Ramat-Gan 5290002, Israel
[2]Institute for Nanotechnology and Advanced Materials, Bar-Ilan University Ramat-Gan 52900, Israel
[†] These authors contributed equally to this work.
[*]Corresponding authors: tomer.lewi@biu.ac.il; doron.naveh@biu.ac.il

## Abstract

Applying direct growth and deposition of optical surfaces holds great promise for the advancement of future nanophotonic technologies. Here, we report on a chemical vapor deposition (CVD) technique for depositing amorphous selenium (a-Se) spheres by desorption of selenium from $Bi_2Se_3$ and re-adsorption on the substrate. We utilize this process to grow scalable, large area Se spheres on several substrates and characterize their Mie-resonant response in the mid-infrared (MIR) spectral range. We demonstrate size-tunable Mie resonances spanning the 2-16 µm spectral range, for single isolated resonators and large area ensembles, respectively. We further demonstrate strong absorption dips of up to 90% in ensembles of particles in a broad MIR range. Finally, we show that ultra-high-Q resonances arise in the case where Se Mie-resonators are coupled to low-loss epsilon-near-zero (ENZ) substrates. These findings demonstrate the enabling potential of amorphous Selenium as a versatile and tunable nanophotonic material that may open up avenues for on-chip MIR spectroscopy, chemical sensing, spectral imaging and large area metasurface fabrication.

## 1. Introduction

Subwavelength Mie resonators are key for achieving light manipulation with low losses and have shown great promise as enabling components of meta-optics and nanophotonic technologies [1,2]. Benefiting from a rich optical response of multipolar magnetic and electric resonances, high-index Mie resonators have enabled new classes of low-loss transmissive nanophotonic devices and metasurface functionalities such as metalenses [3–7], beam deflectors [8–10], holograms [11–13], enhanced nonlinear optical processes [14,15], efficient nanoantennas [1,16], antireflection coatings [17], enhancement of circular dichroism [18], to name a few. Low-loss dielectric nanoparticles form the basic unit cells for high efficiency metasurfaces and have played a pivotal role in nanoantennas [1,2,16], photo-detectors [19,20], nano lasers [21], antireflective coatings [22], thermal emitters [23], nanosphere lithography [24], optical forces [25] among others. Silicon nanoparticles have attracted most attention [2] due to the maturity of fabrication and processing techniques along with CMOS compatibility and electronic integration. However, other materials are required for operation in different spectral ranges (visible, longwave infrared) [3,26], active functionality [27–30] or integration with other material systems [31]. Chalcogenide materials emerge as promising candidates to fill this gap, owing to their extraordinary properties such as high permittivity [32,33], anomalous thermo-optic effects [32], phase transitions [34–36], photo-darkening [37], and lasing [31] , which can be harnessed for advanced photonic technologies. Among the chalcogenides, Selenium (Se) is a highly transparent semiconductor across the entire infrared range and exhibits some extraordinary electrical and optical properties such as high photoconductivity, large birefringence, anisotropic thermo-conductivity, high piezoelectric and thermo-electrical response [38,39]. Se also exhibits negligible optical losses at wavelengths longer than ~ 600nm, where the real part of its refractive index is non-dispersive [40]. So far, studies on Se nanoparticles have focused on the visible and near-infrared ranges [18], including synthetic routes to nanowires [39], nanospheres, and spherical particle colloids [41,42].

Here, we demonstrate for the first time, desorption of selenium from $Bi_2Se_3$ and re-adsorption on the substrate, resulting in condensation of amorphous selenium (a-Se) spheres. While CVD processes involving solid precursors are well studied, this is the first example of combined dissociation and sublimation of the precursor, leading to unique morphological and structural condensation mediated by thermal and compositional equilibrium. We study the Mie-resonant characteristics of a-Se in various morphologies and demonstrate Mie-resonances spanning the entire 2-16 µm MIR range. Using single-particle infrared spectroscopy, we demonstrate size-tunable magnetic and electric resonances in truncated spherical particles, culminating in strong absorption (90%) resonance dips in ensembles of resonant particles. Finally, we show Mie resonance pinning effect in Se resonators placed on an epsilon-near-zero (ENZ) substrate. In the vicinity of the ENZ wavelength, high-Q resonances (Q~40) emerge with the resonance position independent of the resonator size. These findings demonstrate the potential of Se Mie-resonators as a versatile low-loss building block for infrared applications.

## 2. Results and Discussion

Spherical a-Se deposits were achieved by CVD-like growth on various substrates, including Si with 90nm oxide layer, C-plane sapphire, and (100) Si, with typical substrate sizes of 1 - 2 cm$^2$ (see supporting information for full fabrication details). The growth process is schematically illustrated in **Figure 1** (see also a scaled drawing, **Figure S1**). The growth was performed in low-pressure quartz tube with $Bi_2Se_3$ as the starting precursor for the synthesis. The substrates on which a-Se was grown were placed at different distances from the precursor $Bi_2Se_3$. The temperature at the precursor position was 540 and 750 °C, where in all temperatures the preferred sublimation of Se resulted with a deposition of a-Se spheres, with some minor influence on the morphology (sphere size and density). At fixed $N_2$ carrier gas flow conditions (3 sccm), the temperature of the substrate and its distance from the $Bi_2Se_3$ source are dominant and a-Se spheres were grown at the temperatures of 450 and 540°C. The substrate temperature and its distance from the $Bi_2Se_3$ source are controlled by the temperature at the centre of the furnace (see supporting information for details). As substrate temperatures increases, the sphere diameter increases and so is the spacing between spheres. The vaporization of $Bi_2Se_3$ is known to have a kinetically-dependent composition [43], where initially the selenium sublimation dominates the process – corresponding to the reaction $2Bi_2Se_3(s) \rightarrow Bi_4Se_5(s) + \frac{1}{2}Se_2(g)$ with $\Delta H° = 104\text{-}114$ kJ [44]. Interestingly, the deposition of selenium on our substrate is a result of two consecutive phase transitions where the first is into liquid phase (red-marked temperatures in Figure 1b). The vapor condensation rate on the substrate and surface tension of the drop dictates the drop diameter. An additional phase transition, driven by the solute of $Bi_2Se_3$ and Bi vapor – resulting in freezing of the liquid solution corresponding to the observed ~98% Se (**Figure 1b**) [45]. Qualitatively, the size and density of grown solid spheres also depend on flux of Se vapor that is controlled by the $N_2$ flow relative to the precursor to substrate distance and is a scalable method for coating large areas with amorphous Se (a-Se) in spherical morphology, with control over the average size of the spheres.

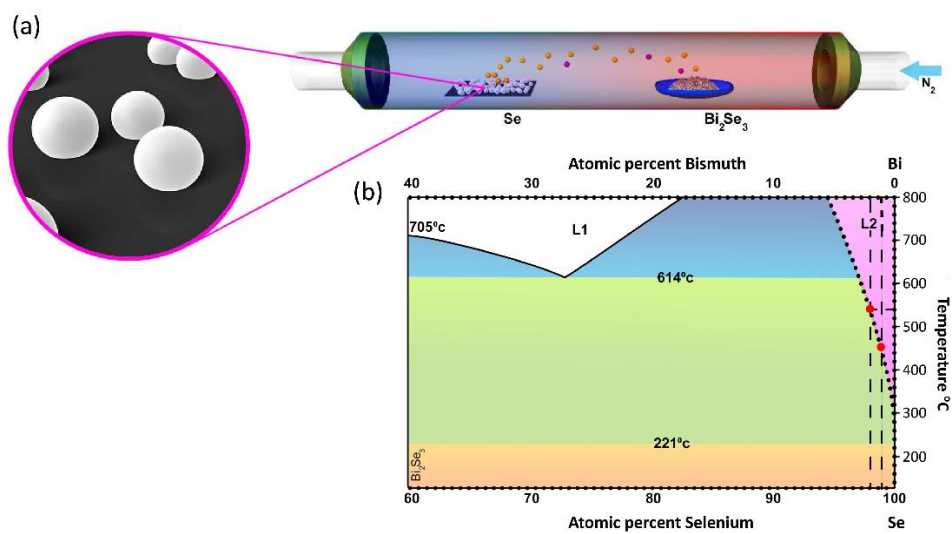

**Figure 1**: CVD growth process (a) Schematic illustration of the a-Se sphere formation by desorption of selenium from $Bi_2Se_3$ and re-adsorption on the substrate. (b) Binary phase diagram of Bi-Se showing the liquid Se (pink area) and the boundary with solid solution (green area).

In the following, we describe the inspection of the grown a-Se as optical resonators in the MIR. First, the fabricated Se spheres were characterized for their morphology, elemental composition, and amorphous phase. Scanning Electron Microscopy (SEM) survey of sub-micron and micron sized spheres were observed to analyze the morphology of the particles, as shown in **Figures 2(a, b)**. SEM micrographs at normal and various tilting angles reveal that the CVD grown resonators are spherical in shape and truncated at the substrate interface, with diameter distribution between 0.6μm and 5μm and aspect ratio of height to diameter (h/D) of ~0.8. SEM survey (a few millimeters is size) shows particle of relative homogeneous size distribution (**Figure 2a**). More details and discussion about size distribution are given in the supporting information (**Figure S2**). **Figure 2(d)** presents a representative energy dispersive X-ray (EDAX) spectroscopy, performed on a single Se sphere on a Sapphire substrate, shown in **Figure 2(b)**. The average composition of ~98% Se in the compound is in good agreement with the expected percentage (**Figure 1b**) and indicates the formation of Selenium resonators. Additional to the composition, the amorphous phase of Bi:Se solid solution was inferred from x-ray diffraction (XRD) on macroscopic ($10 \times 10 \; mm$) samples as presented in **Figure 2(c)**, showing only the peaks of the sapphire substrate (see supporting information for more details).

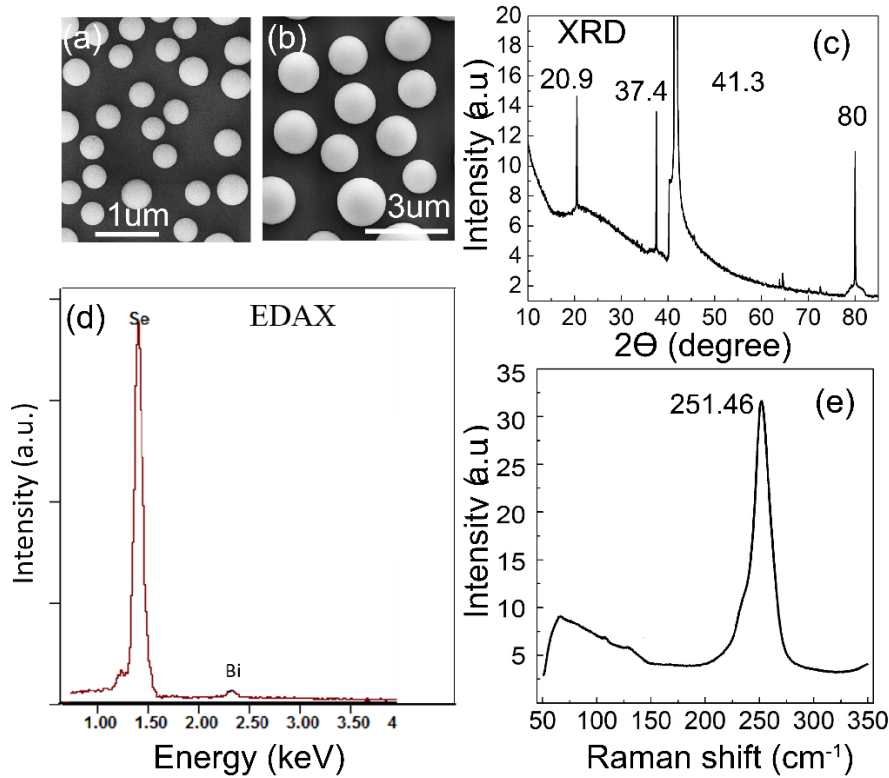

**Figure 2**: (a, b) SEM images of CVD grown sub-micron and micron sized resonators, respectively. (c, d) X-ray diffraction (XRD) and energy dispersive X-ray (EDAX) spectroscopy, respectively. (e) Raman spectrum as collected from a single sphere taken at 633 nm excitation.

Single-particle Raman spectrum shown in **Figure 2(e)** further supports the formation of amorphous phase. The Raman peak at 251.46cm$^{-1}$ is a signature of the stretching of Se-Se mode in the amorphous state [40] [46].

Representative single-particle spectra of an isolated Se resonator are presented in **Figure 3(a)** (inset shows SEM side-view of the resonator at 55° angle). **Figure 3(c)** illustrates the configuration of the optical setup used for the experimental measurement of single-particle Fourier Transform Infrared (FTIR) spectroscopy in reflection mode. The setup includes an FTIR (Nicolet, iS50R) spectrometer coupled to an infrared microscope (Nicolet, Continuum Infrared Microscope). The experimental multipolar Mie scattering spectral response shown is **Figure 3a** exhibits magnetic dipole (MD), magnetic quadrupole (MQ) and electric quadrupole (EQ) resonance modes, as revealed from FDTD numerical calculation (**Figure 3a**, red curve) and from the corresponding electric and magnetic field distributions (Figure 4 a-f). The multipolar scattering response and Mie coefficient decomposition of a perfect spherical resonator of the same diameter, calculated using analytical Mie theory, is presented in **Figure 3b**. Interestingly, both the truncated (**Figure 3a**) and perfect spheres (**Figure 3b**) spectra show that the ED resonance is suppressed in the total scattering response (black curve **Figure 3b**). This is due to the intermediate refractive index value of Se (n=2.65) [47] which results in a significant overlap between the MD and ED mode at the ED resonance wavelength. We note that the spectra of the truncated sphere on the SiO$_2$/Si substrate is slightly red-shifted with respect to the sphere in homogeneous medium (air), with some broadening of resonances which mostly arise from substrate effects [16]. However, the most prominent effect of the substrate is the suppression of the scattering efficiency of the MD mode, resulting in a scattering response dominated by the MQ resonance.

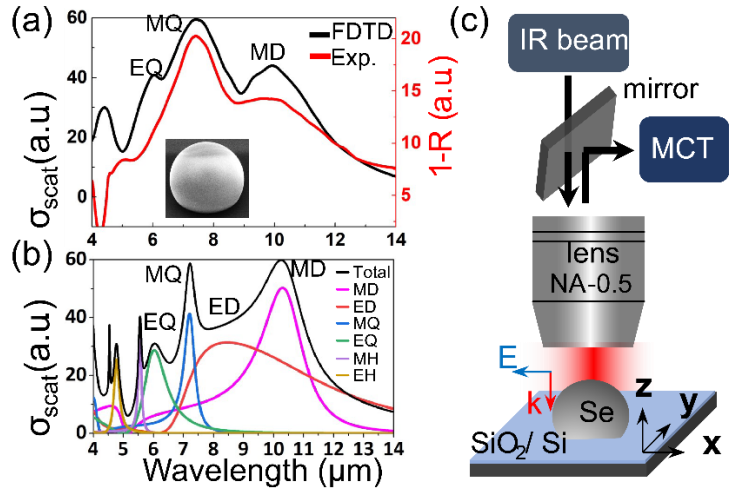

**Figure 3**: Infrared spectra of single Se Mie resonator. (a) Experimentally measured (red) and FDTD numerically calculated (black) infrared spectra of single truncated Se resonator with diameter d= 3.3μm on SiO2/Si substrate (side-view of resonator at 55o angle, shown in inset), respectively. In the FDTD calculations, we used refractive index value of n=2.65 for the a-Se resonator [47] (b) The total calculated Mie scattering spectra (black) of a perfect spherical resonator with d= 3.3μm in air, along with multipolar decomposition of the Mie coefficients (colored lines) of the MD, ED, MQ, EQ, MH, EH (MH=magnetic hexapole, EH=electric hexapole) modes. (c) Illustration of the experimental optical setup configuration.

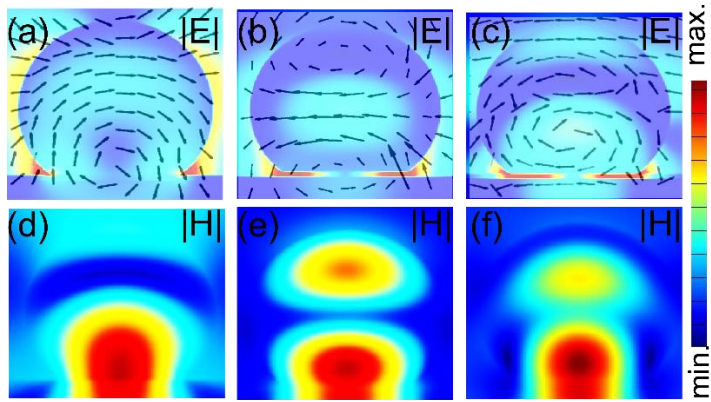

**Figure 4**: Multipolar electric and magnetic field distributions of a single truncated Se resonator with diameter d= 3.3μm on SiO2/Si substrate, presented in Figure 3. (a, b, c) represent the electric field distribution at the MD (9.9 μm), MQ (7.4 μm), and EQ (5.98 μm) resonances, respectively, in the x-z plane. (d, e, f) represents the corresponding magnetic field distribution at the same resonance wavelengths.

The ability to engineer Mie resonance wavelengths with size is fundamental and is necessary for any application. **Figure 5a** and **5b** present numerically calculated FDTD and experimentally measured spectra of particles of variable diameters (with fixed aspect ratio of h/D ~ 0.8), respectively, exhibiting size-dependent multipolar Mie resonances spanning the 2-12 μm spectral range. All FDTD simulations were modeled based on exact size and morphology of the meta-atoms which were extracted from SEM imaging at various tilting angles.

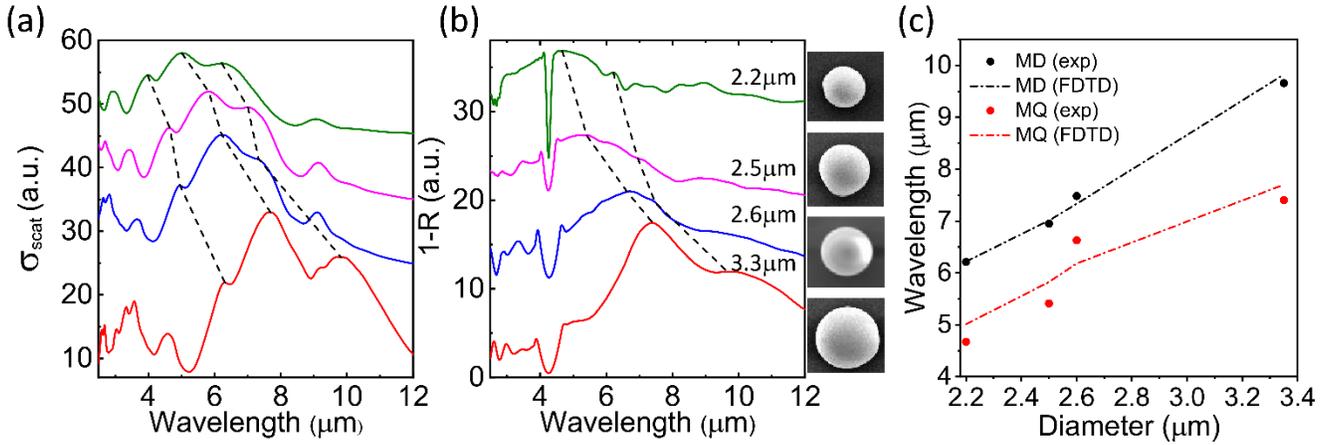

**Figure 5**: Size-dependent tuning of Mie resonances in Se resonators. (a) Simulated FDTD scattering spectra of the experimentally measured particles shown in (b). (b) Experimentally measured spectra of variable sized particles. Inset shows corresponding SEM images. The sharp dips in the experimental spectra at ~4.2μm are due to $CO_2$ absorption in the atmosphere. The black dashed lines in (a,b) are guides to the eye indicating the red-shift of the MD and MQ resonance modes. (c) Extracted size-dependent resonance wavelength shifts of the MD and MQ modes, respectively.

The black dashed line in **Figures 5a** and **5b** represents a guide to the eye for the MD and MQ resonance red-shifts as resonator diameter is increased. **Figure 5(c)** shows the size-dependent FDTD calculated and experimentally extracted MD and MQ resonance wavelength shifts. The experiential observed linear dependence of wavelength with size is in excellent agreement with FDTD predictions and was also observed for Se resonators on various other substrates (e.g. sapphire, Au, Si).

Implementations of size-tunable dielectric Mie resonators in devices and metasurfaces ultimately require easy, scalable, and large-area fabrication methods. For instance, typical Mie resonator-based metasurfaces and devices are fabricated with conventional top-down semiconductor processing techniques. On the other hand, bottom-up chemical synthesis of Mie resonators could provide alternative applications such as, e.g., coatings and paints or self-assembled 3D metamaterials. Next, we demonstrate size-tunable MIR scattering (absorbed in the substrate) resonances of large area ensembles of Se meta-atoms, which can be exploited for large area antireflection coatings, spectral filtering, and photo-detection technology.

The potential of large area CVD grown Se Mie resonators is demonstrated in **Figure 6**. The ensemble measurements presented in **Figure 6a** (ensembles with averaged particle radii of r=1.75μm and r=620nm, respectively) show strong extinction resonances in both experimentally measured spectra and FDTD calculations. Measured and calculated spectra match very well for both ensembles of particles, with the spectral response being dominated by the MQ resonance of the individual particles. In **Figure 6b**, we compare ensemble reflection spectra of widely separated particles and aggregates of particles, which are densely distributed. These measurements reveal whether the scattering response of individual particles that dominate ensembles of isolated particles, are altered in dense aggregates of particles. The spectra exhibit strong extinction resonances (forward scattering into the substrate) of up to 90%, covering a broad MIR spectrum. The spectra of ensembles with widely separated particles (r=620nm and r=1.75µm, blue and black curves) show 50% reflection modulation at the 3µm and 7.1µm resonance dips, respectively, while the dense ensemble shows up to 90% extinction into the substrate. The dashed black lines mark the location of the MQ resonance modes of individual particles which dominate the scattering spectra for both ensembles. The homogeneity of the ensembles is manifested by the invariance of the resonance location for both widely separated and dense particle ensembles. These findings are also in line with previous studies that have demonstrated the strong robustness of Mie-resonators to position and spatial disorder [48]. It should be noted that Se is highly transparent across the whole infrared spectral range [47]; hence the observed extinction resonances are not due to ohmic-losses, but merely radiation losses manifested by the forward scattering of Mie resonances predominantly into the substrate.

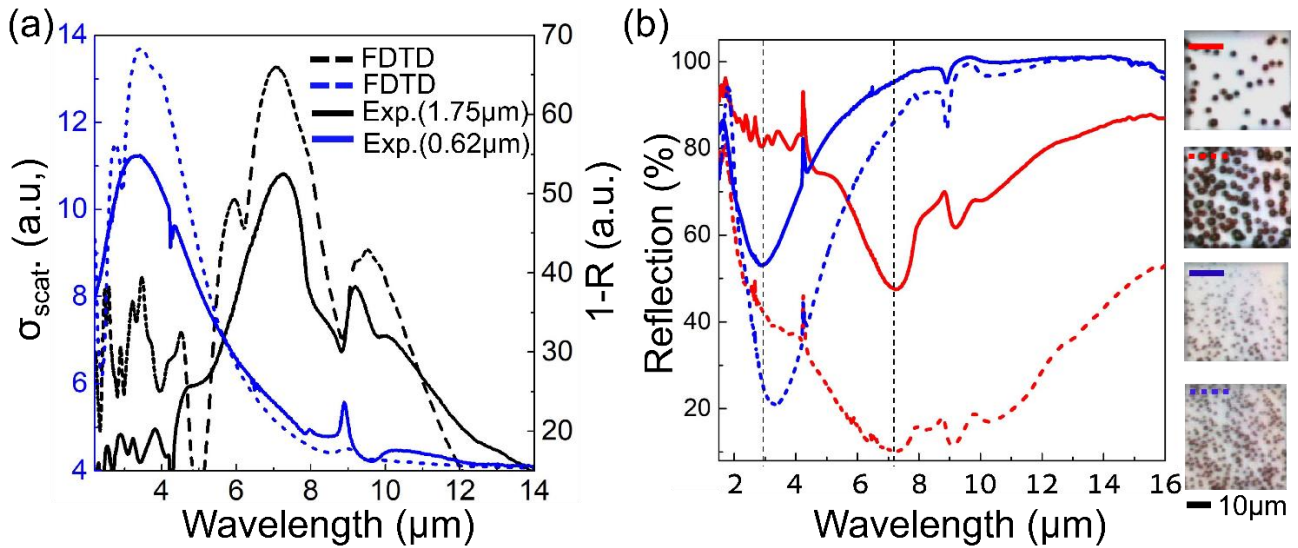

**Figure 6**: Extinction spectra of Se Mie resonator ensembles (a) Experimental (solid line) and FDTD ensemble spectra (dashed lines) with averaged radii of r=620nm (blue) and r=1.75µm (black), respectively. (b) The measured reflection infrared spectra of widely separated and densely packed ensembles of the same averaged size particles r=620nm (solid blue, dashed blue) and r=1.75µm (solid red, dashed red), respectively. Side-inset shows the optical image of the ensembles. Scale bar is 10µm. The particle densities of ensembles of micron size resonators (r=1.75µm) were n=0.03 µm$^{-2}$ (widely separated) and n=0.108 µm$^{-2}$ (densely packed), where for ensembles of sub-micron size resonators (r=620nm), densities were n=0.127 µm$^{-2}$ (separated) and n=0.33 µm$^{-2}$ (densely pack), respectively.

The broadening of the extinction peaks can be attributed to inter-particle cross-talk and is further discussed in the supplementary information (**S4**). Overall, this typical characteristic spectral response, as presented in **Figure 6**, is repeatable for ensembles having the same average radii. The expected scattering spectra will change significantly only for higher order modes, where narrow-linewidth resonances are more sensitive to size/shape variations. These measurements demonstrate the enabling potential of large area Se Mie-resonators in detection, light conversion, thermophotovoltaic cells and energy harnessing applications across the entire infrared range.

Coupling metasurface resonators to ENZ materials has been previously explored due to the unique optical properties arising in the vicinity of a vanishingly small permittivity, including dynamically tunable metasurfaces and wavefront control [49,50] large nonlinearity [51], magnetic field concentration [52] and radiation pattern control [53] to name a few. In the following, we show that the quality factor (Q) of Se resonators can be significantly enhanced when coupled to low-loss ENZ substrates. We demonstrate this by coupling Se resonators to a sapphire substrate, as illustrated in **Figure 7**. Large area Se Mie resonators were CVD-grown on a c-cut sapphire substrate (see supporting information for details). Sapphire exhibits a uniaxial optical character and is transparent from deep-UV up to the beginning of the reststrahlen band (~0.12 eV) where longitudinal optical (LO) and transverse optical (TO) modes induce an ENZ region. The permittivity of Sapphire (ordinary axis) is plotted in **Figure 7c**. The ENZ region at~11 µm is followed by a spectral range of negative permittivity with low losses, which can be used to excite surface phonon polaritons (SPhPs) [54–56] [53]. The sapphire ENZ value at 11.01µm wavelength ($\varepsilon = 0.001 + i0.135$), is marked by black dashed line in **Figure 7(c)**. The low imaginary value indicates that the substrate exhibits low losses at the ENZ wavelength. Reflection infrared spectrum of single isolated Se meta-atom of diameter d=3.49µm (aspect ratio h/d=0.75, inset shows SEM image) on a sapphire substrate along with FDTD calculation are shown in **Figure 7(a)**. The green curve represents FDTD scattering spectrum of an identical sized Se resonator surrounded by air ($\varepsilon$ = 1). Compared to the resonator in air, the Se resonator placed on sapphire exhibits a large red shift. This red shift is due to the negative permittivity of the sapphire substrate at these wavelengths, inducing an interaction with mirror image in the Se particle, thus red shifting the resonances to longer wavelengths [32,57]. The two main features in the experimental spectra of **Figure 7(a)** are on-resonance high-Q spectral peaks of MD mode (Q ~ 8.8) and a localized surface phonon resonance (Q ~ 38) observed at 11.3µm and 14.8µm, respectively (black curve). These values represent a 5-fold increase in quality factors compared to the same resonators on Si substrates. Numerically calculated spectra of the same single particle (red curve in **Figure 7a**) reveal very similar quality factors of Q~10.2 at λ=11.4µm and Q~43 at λ=15.4µm for the two resonances, respectively.

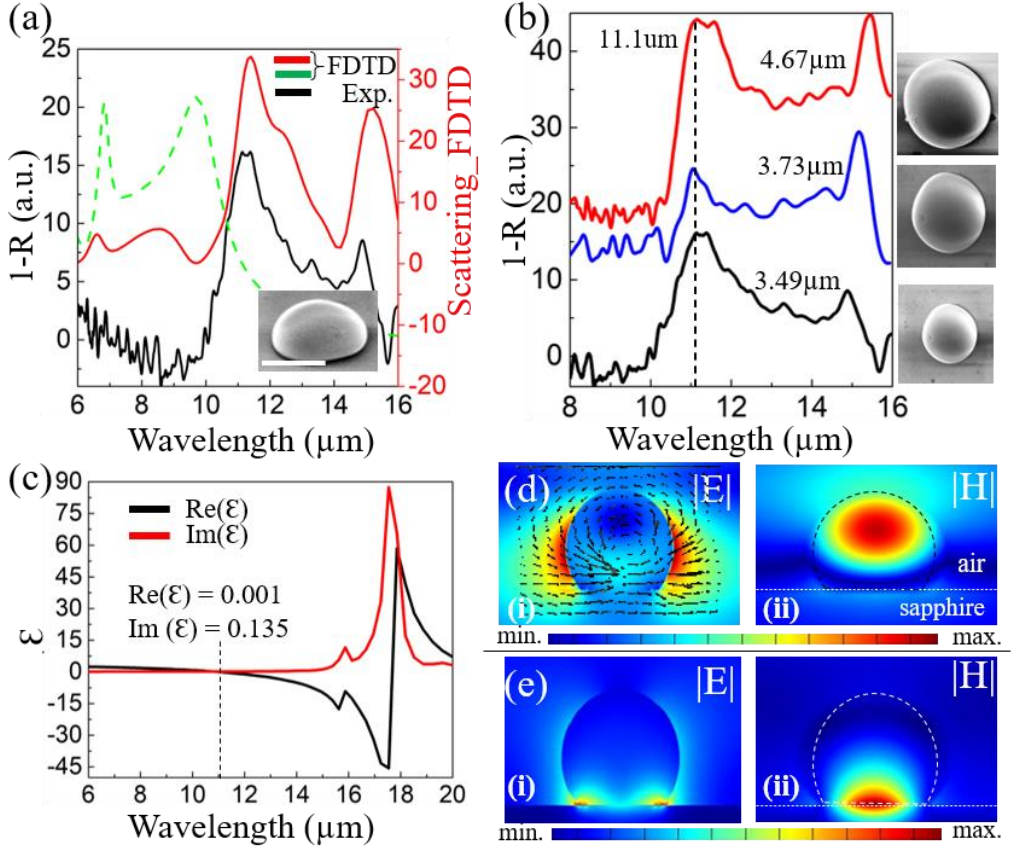

**Figure 7**: Spectral properties of Se resonators on low-loss ENZ substrate. (a) Experimental (black) and numerically calculated FDTD (red) spectra of an isolated single Se resonator (d=3.49μm) on a sapphire substrate. The FDTD calculated spectra of the same Se resonator without the substrate (i.e. in air), is also shown (green curve). Inset shows SEM image of the particle at 55 deg tilt (scale bar is 2μm). (b) Single particle spectra of three different sized resonators on sapphire, exhibiting size-independent resonances at the ENZ wavelength (λ~11.1μm). The ENZ wavelength is marked by a black dashed line manifesting the pinning effect of the resonance. Insert shows SEM images of the particles. The scale bar is 3μm in (a). (c) The real and imaginary parts of permittivity of sapphire calculated from [58]. The ENZ wavelength is marked by a black dashed line. (d, e) Electric and magnetic field distributions at the two spectral features shown in (b). (d) Electric and magnetic field distributions of an MD mode at λ=11.1μm (e) Electric and magnetic field distribution at λ=15.4μm

**Figure 7(b)** presents the spectra of three particles of variable sizes (d=3.49μm, 3.73 μm, d=4.67μm, respectively). The MD resonance peak in all three particles is size-independent and pinned at λ=11.1μm (dashed line in **Figure 7b**). This value is determined by the location of the ENZ in the sapphire and not by the resonator size. This 'pinning' effect of resonance wavelength to the ENZ value can be attributed to the effective index ($n_{eff}$) of the local environment [50], which is significantly dominated by ENZ of the substrate. Such an effect has been previously observed mainly in plasmonic resonators [53]. MD resonance Q-factors as high as $Q_{MD}$ ~19.8 are observed for the d=3.73 μm particle. These MD Q-factors values are higher than previously reported Q-factors in individual meta-atoms in the MIR range, obtained from Si ($Q_{MD}$~7) [16] , Ge ($Q_{MD}$~7) [16] or even from high index PbTe (n~5.7) resonators ($Q_{MD}$~10.8) [32].

The long wavelength peak (at $\lambda \sim 15 \mu m$) of the spectra is due to a localized surface phonon mode. This resonance is slightly red shifted as size increases and is also influenced by the rapid change in the permittivity at these long wavelengths. This resonance is also characterized by narrow linewidths with quality factors reaching Q~43.

To further investigate these spectral features, we used FDTD numerical calculations to study the field distributions. **Figures 7d** and **7e** represent the electric (i) and magnetic (ii) field distributions of these two resonances at $11.1 \mu m$ and $14.8 \mu m$, respectively. The resonance at ~ $11.1 \mu m$ is driven by the ENZ region. Since the resonator is placed on a vanishingly small permittivity, the MD-like mode is concentrated at the top of the resonator and at the resonator-air interface where the denser permittivity (air, $\varepsilon = 1$) exists, and hence the local density of optical states is higher. Furthermore, the low permittivity of the substrate is also responsible for the predominantly backscattered radiation towards the denser medium (air). At longer wavelengths however ($\lambda > 12 \mu m$), the negative permittivity of the Sapphire substrate can support surface phonon modes. Indeed, as can be seen in **Figure 7e**, the spectral feature at $14.8 \mu m$ is a localized surface phonon mode where the fields are concentrated at the resonator-substrate interface. The high Q-factor of this resonance mode (Q~43) is enabled by the relatively low imaginary part of the permittivity in the sapphire substrate. Further analysis also demonstrates that the Se resonator effectively couples the free space radiation of the incoming beam to confined SPhP oscillations at $\lambda = 11.9$ (where $\varepsilon = -1$, see supporting information Figure S5). Altogether this platform achieves extremely high-Q values in single resonators with a moderate refractive index (n=2.65).

### 3. Conclusion

We have demonstrated a novel growth method for a-Se spherical particles on various substrates, and their optical characterization as Mie-resonators. Numerical and experimental observations reveal that the multipolar Mie scattering infrared spectra is dominated by the MD and MQ resonance modes. We demonstrate size-engineered resonances that span the entire MIR range, culminating in strong extinction resonance dips of up to 90% in ensembles of resonant particles. Finally, we demonstrate that high-Q resonances (Q~40) arise when the resonators are placed on low-loss ENZ substrates. These findings highlight the potential and versatility of a-Se Mie resonators and suggest new possibilities for applications such as Mie-resonant paints, coatings, on-chip chemical and biological sensing, spectral imaging and photodetection technology with Selenium resonators.

**Supporting information**

CVD Growth process for a-Se resonators, optical setup, experimental procedures and data acquisition, 3D FDTD simulation methods, electric and magnetic field distribution for a-Se resonators on Si and $SiO_2$ substrates, FDTD spectra for ensembles of different particle densities, electric field distribution of resonators on ENZ and the excitation of SPhP oscillations

**Acknowledgement**

TL would like to thank the Israel Science Foundation for funding this work under grant No. 2110/19; DN would like to thank the European Research Council for the generous funding of this work under the H2020 FET OPEN grant No. 801389.

**Conflict of Interest:**

Authors declares no conflict of interest

# Supporting Information

# Amorphous Selenium Mie Resonators for Infrared Meta-Optics

*Danveer Singh[1†], Michal Poplinger[1†], Avraham Twitto[1], Rafi Snitkoff[1], Pilkhaz Nanikashvili[1], Ori Azolay[1], Adi Levi[1], Chen Stern[1], Gili Cohen Taguri[2], Asaf Albo[1,2], Doron Naveh[1,2*] and Tomer Lewi[1,2*]*

[1]*Faculty of Engineering, Bar-Ilan University, Ramat-Gan 5290002, Israel*
[2]*Institute for Nanotechnology and Advanced Materials, Bar-Ilan University Ramat-Gan 52900, Israel*
[†] These authors contributed equally to this work.
[*]Corresponding authors:  tomer.lewi@biu.ac.il; doron.naveh@biu.ac.il

## 1. CVD Growth process for amorphous Se resonators

**1.1 Substrate pre-clean process:** Amorphous Se resonators were grown on three different types of substrates: on Sapphire [0001], Si/SiO$_2$ (with 90 nm thin film of SiO$_2$) and Si. Before starting the CVD growth process, the surfaces of the substrate were gone through the following pre-deposition cleaning steps. First, samples were immersed in an ultrasonic bath with acetone and isopropyl alcohol solvents for 5 minutes to remove contaminants. Afterwards, the samples were cleaned in NMP (*N*-Methyl-2-Pyrrolidone) at 80ºC for 15 min. To remove organic leftovers on the surface of the substrate, the substrates were treated in a sulfuric peroxide solution (H$_2$SO$_4$:H$_2$O$_2$) in a ratio of 3:1 at 90ºC for 15 min, followed by washing in deionized water.  The substrates were then entered into an oxygen plasma for 5 minutes at a power of 150W.

### 1.2 CVD process

The synthesis of amorphous Se was carried out by the CVD method in a tube furnace (Carbolite MTF 12/25/250**)** in a quartz tube of 2.16cm inner diameter. The second and smaller concentric quartz trimmed tube with a diameter of 1.57cm and total tube length of 30 cm was inserted into the larger tube (**Figure S1**). In the smaller tube, the precursor was placed in a quartz boat. The starting precursor for the CVD synthesis was Bi$_2$Se$_3$ (Alfa Aesar – 99.999%) and it was placed at varying positions relative to the center of the furnace: (i) the temperature of both the Bi$_2$Se$_3$ source and substrate were kept at 540 ℃; (ii) the source was kept at 750 ℃ and the substrate at 450 ℃. In both cases the sublimation of Se was manifested by re-condensation of a-Se. However, the condensation of condition (i) manifested in large and sparse spheres, while at condition (ii) the a-Se deposition resulted with the morphology of **Figures 1**, **2** of the main text. **Figure S1** displays the experimental setup.

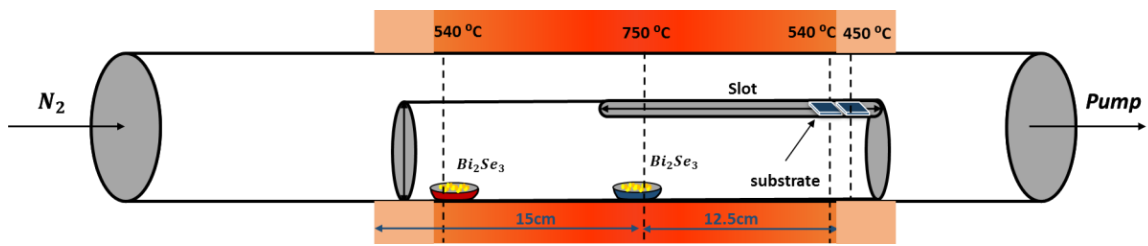

**S1:** The experimental setup of a-Se CVD: (i) both substrate and source kept at 540 ℃; (ii) source at 750 ℃ and substrate at 450 ℃. The heat zone of the furnace (25 cm long) is insulated with a 2.5 cm ceramic from both its sides.

The substrates on which amorphous Se was grown were placed on top of the slot with the surface oriented towards the downside (face down).

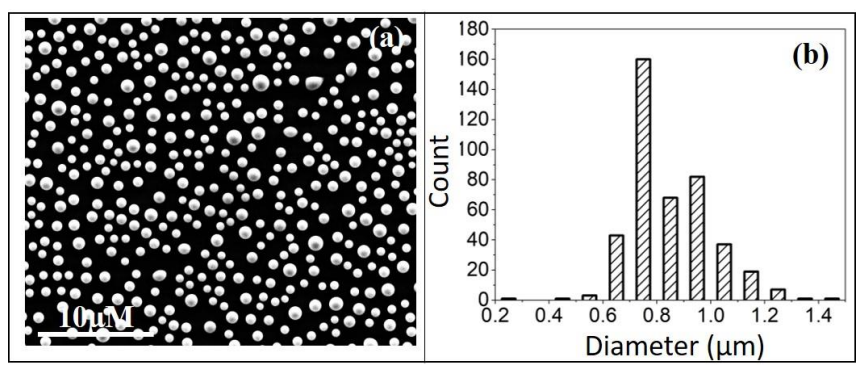

**S2:** Size distribution analysis. (a) represents the SEM image of CVD grown Se resonators with different particle sizes. (b) indicates the histogram of particle's size distribution.

The grown resonators have size distribution ranging from 0.6µm to 5µm. **Figure S2(a)** shows SEM image of a typical area containing Se resonators with relatively uniform size distribution (mean size 0.76±0.12µm) indicated in the histogram **S2(b)**. The size distribution was calculated using ImageJ software.

## 2. Optical setup

In our experiments, we used a setup consisting of a FTIR spectrometer (Nicolet, iS50R) coupled to an infrared microscope (Nicolet, Continuum Infrared Microscope). Spectra were acquired with reflection measurements, where infrared radiation was focused on the sample by a 20x Cassegrain objective with NA=0.5, defining a beam spot size of ≈250µm. The reflected light was collected by the same 20x objective. A variable knife-edge square aperture located in the image plane was used to define the signal collection area and only the signal originating from that area was directed to the MCT detector. For ensemble measurements, we used aperture size of 30µm X 30µm, whereas for single particle measurements the aperture size was optimized to maximize the signal to noise ratio (typically it was 20µm x 20µm). All FTIR spectra were gathered as $P_{sample}/P_{background}$, where $P_{sample}$ is collected from an area with the particles present and $P_{background}$ is collected from an adjacent area with no particles present.

## 3. 3D FDTD Simulation Methods

We have used a commercial three-dimensional FDTD software, Lumerical's Solutions [1] to calculate the scattering spectra of single isolated Se meta-atoms, as well as ensembles of particles, placed on $Al_2O_3$/Si/$SiO_2$ substrates. We have used a non-uniform three-dimensional square-shaped mesh with a minimum size of 0.2nm and a uniform 20nm overriding mesh. This configuration was used to obtain higher resolution at the sharp geometrical discontinuities present at the sides and in-between the particle and substrate. A perfectly matching layer (PML) was used to eliminate the unwanted reflections at the boundaries and was placed sufficiently far from the particle (around 10um) on each side. To model the substrate materials, we have used frequency-dependent dielectric optical constants of Si and $SiO_2$ from Palik [2] and the optical constants of $Al_2O_3$ from [3]. In the simulation, the Total-Field Scattered-Field (TFSF) source was used to illuminate the resonators. The observation of scattering spectrum and the magnitude of electric and magnetic near fields was calculated using two-dimensional monitors that are placed in the middle of the particle.

For the numerical calculations, we used the refractive index (n) value of Se n = 2.65 for all the wavelengths from 2µm to 16µm [3]. The particle size and shape were extracted from SEM images at various tilting angles.

## 4. Electric and magnetic field distribution for a-Se resonators on Si and $SiO_2$ substrates

FDTD simulation were performed to investigate the electric and magnetic field distributions at the Mie resonance wavelengths with two different substrates. The large local density of optical states (LDOS) of the Si substrate, pulls the optical modes of the truncated sphere downwards to the interface with SiO2/Si Substrate. However, beyond the leakage to the substrate, the field distributions are quite similar. The effect of the SiO$_2$ is negligible and similar field plots are obtained for a pure Si substrate.

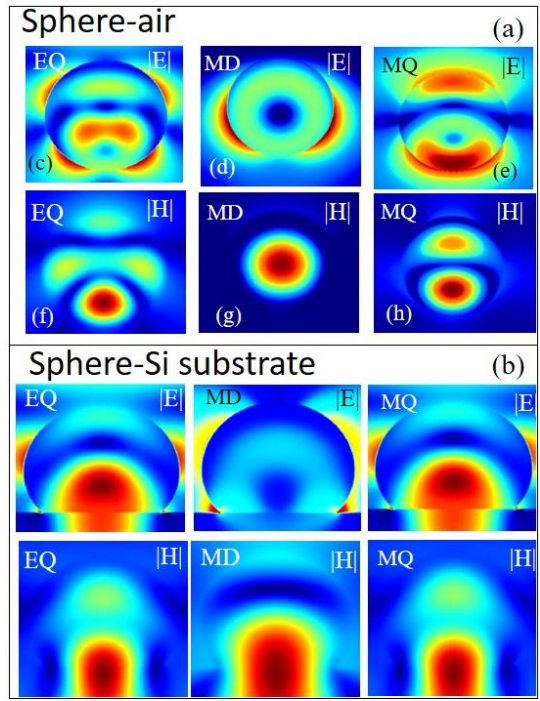

**S3:** Electric and magnetic field distribution of Se particle (d=3.3µm) placed on the Si substrates, and in homogenous medium (air), respectively.

**5. FDTD spectra for ensembles of different particle densities**

**6.**

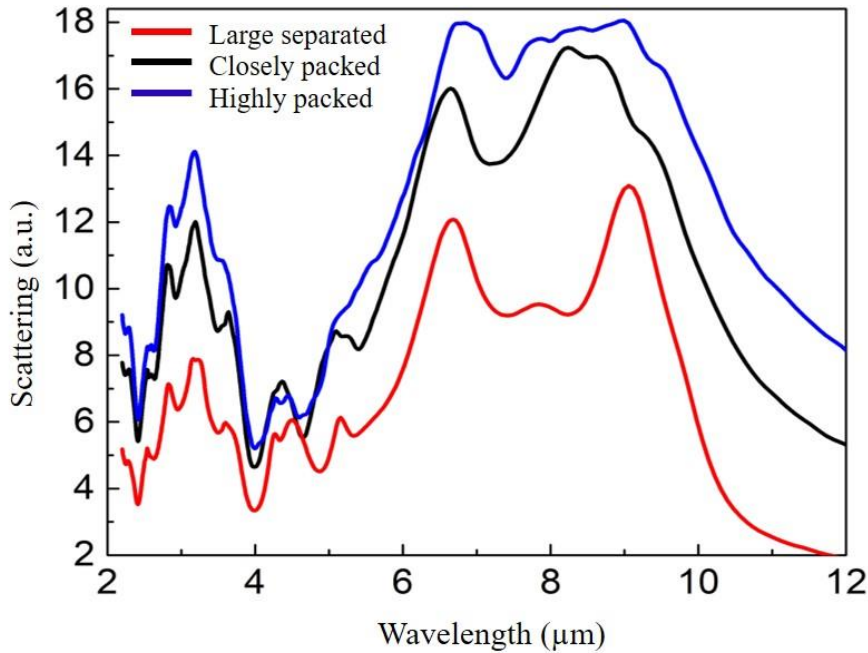

**S4.** The scattering spectra of three different configurations of ensembles of Se resonators. The broadening of spectrum can be observed as the particles are closely packed with each other.

We simulate various ensemble configurations, similar to the experimentally observed ensembles presented in **Figure 6** of the main text. In this context, the scattering spectrum was analyzed for the various ensemble configurations which varied from widely dispersed to tightly packed particle ensembles. Two main trends are observed in the spectra: i) Denser ensembles scatter light more efficiently- a result of the larger area covered with resonators. ii) The spectra of denser ensembles are broadened which can be attributed to the inter particle cross-talk.

## 7. Electric field distribution of resonators on ENZ and the excitation of SPhP oscillations

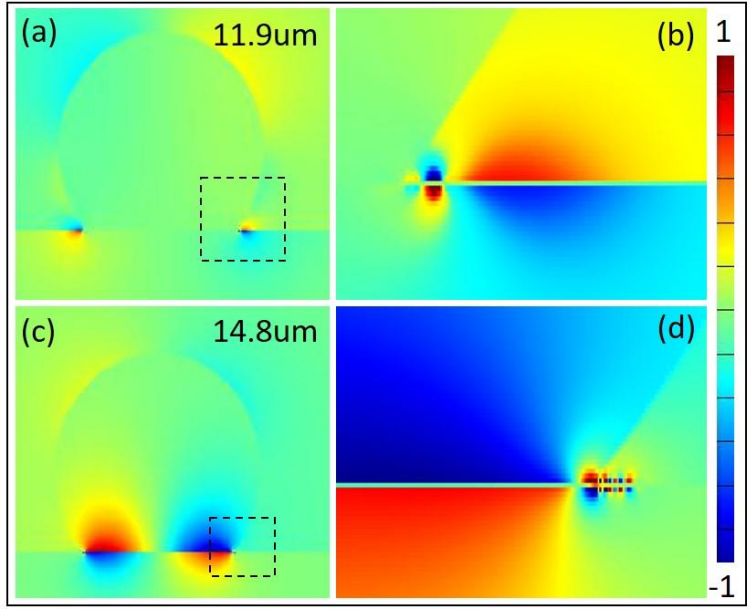

**S5:** Analysis of electric field distribution at λ=11.9μm of Se resonator on sapphire as shown in Figure 7 of main text. At λ=11.9μm, the real part of the substrate permittivity is ε =−1. (a) represents the electric field distribution in x-z plane at λ=11.9um. When ε =−1, the system supports the excitation of surface phonon polaritons (SPhP). The field is confined outside and at the geometrical discontinuity of resonator and substrate, with some surface oscillations pointing outwards, as can be seen in (b). (b) zoom-in on the square dashed area indicated in (a). (c) represents the electric field distribution at the 14.8 μm resonance, where the maximum field is confined within the resonator, as can be seen in the zoom-in area indicated by the dashed square. (d) Zoom-in on the square dashed area marked in (c).